\documentclass[twocolumn,pra,aps]{revtex4}
\usepackage[dvips]{graphicx}
\usepackage{amsmath}
\usepackage{amsfonts}
\usepackage{amssymb}

\draft

\newcommand{\ar}{\begin{eqnarray}}
\newcommand{\br}{\end{eqnarray}}
\newcommand{\ica}{I_{\rm Ca}}

\newcommand{\naca}{I_{\rm NaCa}}

\begin{document}

\title{Coupled Dynamics of Voltage and Calcium in Paced Cardiac Cells}
\author{Yohannes Shiferaw$^{1,2}$, Daisuke Sato$^{1}$, and Alain Karma$^{1}$}
\affiliation {$^{1}$Department of Physics and Center for
Interdisciplinary Research on
Complex Systems,\\
Northeastern University, Boston, MA 02115}
\affiliation{$^{2}$Department of Medicine and Cardiology,
University of California, Los Angeles, CA 90095-1679}

\date{\today}

\begin{abstract}
We investigate numerically and analytically
the coupled dynamics of transmembrane voltage and intracellular
calcium cycling in paced cardiac cells using a detailed physiological
model and its reduction to a three-dimensional discrete map.
The results provide a theoretical framework to interpret various
experimentally observed modes of instability ranging
from electromechanically concordant and discordant alternans
to quasiperiodic oscillations of voltage and calcium.
\end{abstract}

\maketitle

\begin{figure}
\includegraphics[width=6.50cm]
{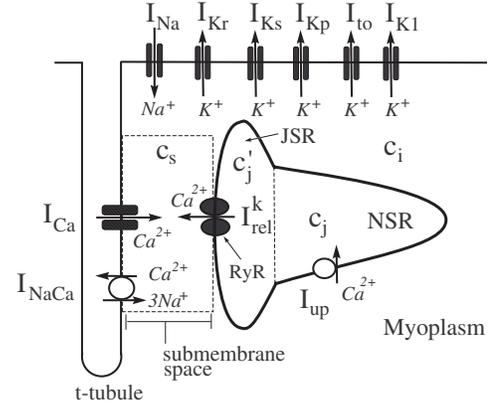} \caption{Illustration of currents that control the
dynamics of voltage and intracellular calcium cycling. }
\label{calcium-model}
\end{figure}

Over the last decade, there has been a growing recognition that
dynamic instability of the cardiac action potential can play a
crucial role in the initiation of life-threatening arrhythmias
\cite{karma,garfinkel,modalt,expalt,hall,foxmap}.  Most studies to
date have focused on the dynamics of the transmembrane voltage
governed by the standard equation \ar \dot V=-(I_{\rm ion}+I_{\rm
ext})/C_m,\label{voltage} \br where $C_m$ is the membrane
capacitance, $I_{\rm ion}$ is the total membrane current, which is
the sum of the individual currents for Na$^+$, K$^+$, and
Ca$^{2+}$ ions depicted schematically in Fig. \ref{calcium-model},
and $I_{\rm ext}$ is the external current representing a sequence
of suprathreshold stimuli equally spaced in time by $T$, the
pacing period. A widely used approach to model the nonlinear
dynamics of voltage is the one-dimensional discrete map
$A_{n+1}=f(T-A_n)$ which relates the action potential duration
(APD) at two subsequent beats via the restitution curve,
$A_{n+1}=f(D_n)$, where $D_n$ is the interval between the end of
the previous action potential and
the next \cite{karma,garfinkel,modalt,expalt,hall,foxmap}. The periodic fixed 
point of this map corresponding to the stable 1:1 rhythm undergoes
a period-doubling instability to alternans, a sequence LSLS... of
long (L) and short (S) APD, when the slope of the restitution
curve is $>1$.

Even though this map has been successful to model the unstable
dynamics of voltage in some ionic models \cite{modalt} and
experiments \cite{expalt}, its predictions are inconsistent with a
wide range of observations \cite{hall,foxmap,pruvot,lab}. For
example, Hall \emph {et al.} \cite{hall} found that alternans can
be absent even when the slope of the restitution curve is
significantly larger than one, and conversely alternans are
observed under ischemic conditions in which the restitution curve
is flat \cite{lab}.  An important limitation of the restitution
relationship is highlighted by recent experimental
\cite{pruvot,chudin,eisner-sr} and theoretical studies
\cite{shiferaw} which suggest that alternans may result from an
instability of intracellular calcium cycling. The coupled
nonlinear dynamics of voltage and calcium, however, remains
largely unexplored.

In this letter, we investigate this dynamics by a numerical
study of a detailed physiological model and an analysis of
the dynamics based on iterated maps. The model consists of Eq.
\ref{voltage}, with membrane currents (Fig. \ref{calcium-model})
modeled based on modifications by Fox {\it et al.} \cite{fox} of
the Luo-Rudy currents \cite{rudy}, coupled to equations from a recent
model of calcium cycling \cite{shiferaw}
\begin{eqnarray}
& &\dot c_s=\frac{\beta_s v_i}{v_s}\left[
I_{\rm rel}-\frac{c_s-c_i}{\tau_s}-\ica
+\naca\right] -\beta_s I^s_{\rm tr},~~~~~~\label{c1}\\
& &\dot c_i=\beta_i\left[\frac{c_i-c_s}{\tau_s} -I_{\rm
up}-I^i_{\rm tr}
\right], \label{c2}\\
& &\dot c_j=-I_{\rm rel}+I_{\rm up},\label{c3} \\
& &\dot c_j'=\frac{c_j-c_j'}{\tau_a}, \label{c4}\\
& &\dot I_{\rm rel}=g\ica Q(c_j')-I_{\rm rel}/\tau_r,\label{c5}
\end{eqnarray}
where $c_s$, $c_i$, and $c_j$ are the concentrations of free
Ca$^{2+}$in a thin layer just below the cell membrane (submembrane
space), in the bulk myoplasm, and the sarcoplasmic recticulum
(SR), with volumes $v_s$, $v_i$, and $v_{\rm sr}$, respectively,
where the SR volume includes both the junctional SR (JSR) and the
network SR (NSR); $c_j'$ is the average JSR concentration in the
whole cell as defined in Ref. \cite{shiferaw}. The concentrations
$c_s$ and $c_i$ are in units of $\mu$M, whereas $c_j$ and $c_j'$
are in units of $\mu$M$v_{\rm sr}/v_i$. All Ca$^{2+}$ fluxes are
divided by $v_i$ and have units of $\mu$M/s. Instantaneous
buffering of calcium to SR and calmodulin sites in $v_i$ and $v_s$
is accounted for by the functions $\beta_s\equiv \beta(c_s)$ and
$\beta_i\equiv \beta(c_i)$, and the currents $I^{s,i}_{\rm tr}$
describe time-dependent buffering to troponin C \cite{shiferaw}.

Calcium release from the SR is triggered by calcium entry into the
cell via calcium-induced-calcium-release (CICR) \cite{fabiato}.
Release occurs at a very large number of junctions where several
L-type Ca channels ($\ica$) and a few release channels (ryanodine
receptors; RyRs) face each other in close proximity. Only one of
these junctions is shown in Fig. \ref{calcium-model} for clarity.
The total release current for the whole cell is the sum $I_{\rm
rel}=\sum_{k=1}^{N(t)} I^{k}_{\rm rel}$, of local currents
$I^{k}_{\rm rel}$ at each junction where release channels are
activated. Active junctions appear as bright localized spots, or
``sparks'', in confocal microscope imaging of calcium activity
\cite{bers}. The number of sparks $N(t)$ varies in time since
sparks are recruited stochastically and extinguish. The model
takes into account this spatially localized nature of release and
the dynamical equation for the release current (Eq. \ref{c5})
captures phenomenologically three key experimental observations:
(i) sparks are recruited at a rate proportional to the whole cell
$\ica$, or $\dot N\sim \ica$ \cite{collier}, which insures that
calcium release is graded with respect to calcium entry
\cite{bers, wier}, (ii) the spark life-time $\tau_r$ is
approximately constant, and (iii) the amount of calcium released
increases with SR concentration (SR-load) \cite{shannon}.

{\it Instability mechanisms}. Ca$^{2+}$ alternans, a
period-doubling  sequence $lsls$... of large ($l$) and small ($s$)
calcium transient ($lsls$ peak $c_i$), can occur independently of
voltage alternans in experiments with a single cell paced with a
periodic voltage waveform \cite{chudin}. Both theoretical analyses
\cite{shiferaw,eisner} and recent experiments \cite{eisner-sr}
support that a steep dependence of release on SR-load is the
underlying mechanism of these alternans. The sensitivity of
release to SR-load is controlled in the model by the slope of the
function $Q(c_j')$ at high load
\begin{equation}
u\equiv dQ/dc_j'.
\end{equation}
For a large enough slope,
the model produces Ca$^{2+}$ alternans
when paced with a periodic voltage waveform \cite{shiferaw}
as in the experiments of Ref. \cite{chudin}.

Steep APD-restitution in the absence
of Ca$^{2+}$ alternans can also induce
APD alternans. This steepness is especially sensitive
to the recovery from inactivation of the
calcium current \cite{fox,rudy}
\begin{equation}
\ica=d \ f \ f_{\rm Ca} \ i_{\rm Ca},
\end{equation}
where $i_{\rm Ca}$ is the single channel current and $d$ ($f$) is
a fast (slow) voltage-dependent activation (inactivation) gate.
For the intermediate range of pacing rates studied in the present
work, increasing the time constant $\tau_f$ of the $f$ gate in the
equation $\dot f=(f_\infty(V)-f)/\tau_f$ steepens APD-restitution
and promotes voltage alternans.

\begin{figure}
\includegraphics[width=5.50cm]
{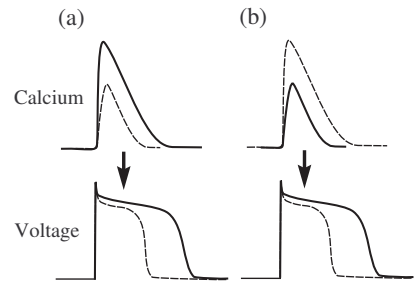} \caption{Illustration of the effect of
an increase in the magnitude of the calcium transient,
which can prolong or shorten the APD for (a) positive
and (b) negative coupling, respectively.
The sign of the coupling depends on the relative
contributions of $\ica$ and $\naca$ to the APD.
The solid or dashed lines correspond to the same beat.}
\label{coupling}
\end{figure}

{\it Voltage-calcium coupling}.
The mutual influence of voltage and calcium during
the action potential is controlled by the membrane currents that
depend on intracellular calcium concentration. These
include $\ica$ and the sodium-calcium exchanger $\naca$.
A crucial property is that a change in the magnitude of
the calcium transient has opposite effects on these
currents with respect to prolonging or shortening
the APD. A larger calcium transient following a larger release enhances
inactivation of $\ica$ via the calcium-dependent
gate $f_{\rm Ca}$, and hence shortens the APD,
but increases the chemical driving force
for Ca$^{2+}$ extrusion from the cell via the exchanger.
Since 3 Na$^+$ enter the cell for every Ca$^{2+}$ extruded,
this increase in driving force increases
the inward membrane current which prolongs the APD.
Therefore, depending on the relative
contributions of $\ica$ and $\naca$, increasing the magnitude of
the calcium transient can either
prolong (positive coupling) or shorten
(negative coupling) the APD, as illustrated in Fig. \ref{coupling}.
The sign of this coupling can be changed
in the model by varying the exponent $\gamma$
in the phenomenological expression
$f_{\rm Ca}^{\infty}=1/\left[1+(c_s/\tilde{c}_s)^\gamma\right]$
for the steady-state value of $f_{\rm Ca}$,
where the constant $\tilde{c}_s$ sets the
concentration range for inactivation.
Increasing $\gamma$ enhances calcium-dependent
inactivation of $\ica$ and tends to
make the coupling negative.

{\it Numerical results}. The dynamics of the model was studied
numerically as a function of the two instability parameters $u$
and $\tau_f$ which promote Ca$^{2+}$ and voltage alternans,
respectively, and for two values of $\gamma$ that were found to
yield a positive ($\gamma=0.7$) and a negative ($\gamma=1.5$)
coupling between voltage and calcium. All the other parameters are
the same as in Ref. \cite{shiferaw,fox} and the pacing period is
fixed to $T=300$ ms.

\begin{figure}
\includegraphics[width=7.50cm]{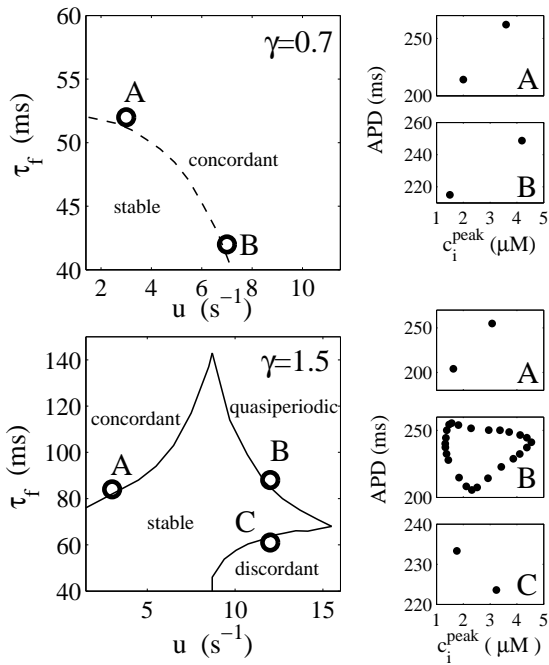}
\caption{Stability boundaries
in the ionic model for positive (dashed line; $\gamma=0.7$) and
negative (solid line; $\gamma=1.5$) coupling. $T=300$ ms.
Examples of steady-state dynamics close to the stability
boundaries are illustrated by plots of peak calcium
concentration ($c_i^{\rm peak}$) vs. APD for a few labelled points.
Higher order periodicities and irregular dynamics are
observed further away from these boundaries.}
\label{stab}
\end{figure}

The results plotted in Fig. \ref{stab} highlight the crucial role
of the coupling between voltage and calcium in the dynamics. For
positive coupling, the instability of the 1:1 periodic state
always occurs through a period-doubling bifurcation to
electromechanically concordant alternans with the long (short) APD
corresponding to a large (small) calcium transient, independently
of whether voltage or calcium is the dominant instability
mechanism. In contrast, for negative coupling, three distinct
modes of instability are found that correspond to (i) concordant
alternans, as for positive coupling, but only when the instability
is dominated by voltage (large $\tau_f$ and small $u$), (ii)
discordant alternans with the long (short) APD corresponding to a
small (large) calcium transient when the instability is dominated
by calcium (small $\tau_f$ and large $u$), and (iii) quasiperiodic
oscillations of APD and calcium transient amplitude with a phase
and a Hopf frequency that vary with $\tau_f$ and $u$ for the in
between case where the instability is driven by both voltage and
calcium. Both electromechanically concordant and discordant
alternans have been widely observed experimentally under various
conditions \cite{discordant-papers}. In addition, there is
experimental evidence for quasiperiodicity in recordings of
voltage \cite{gilmour1} and, more recently, calcium
\cite{entcheva}.

\begin{figure}
\vskip -2 cm
\includegraphics[width=6.0cm]
{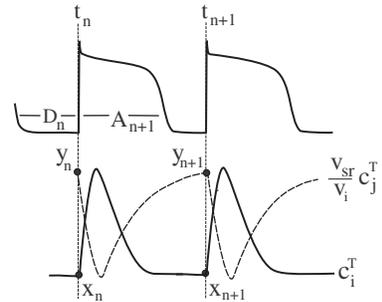} \caption{Definition of map variables.}
\label{map}
\end{figure}

{\it Iterated map of voltage-calcium dynamics}. To interpret our
results, we extend the two-dimensional iterated map developed in
Ref. \cite{shiferaw} for calcium cycling when the cell is paced
with a fixed periodic voltage waveform, to the present case where
the voltage is unclamped. To a good approximation, $c_s\approx
c_i$ and $c_j'\approx c_j$ preceding a stimulus \cite{shiferaw},
such that we only need to track beat-to-beat changes of $c_i$ and
$c_j$. Furthermore, we assume for simplicity that buffering of
calcium is instantaneous such that there exists a unique nonlinear
relationship between the concentration of free calcium $c_i$
($c_j$) and total calcium (free plus bound) $c_i^T$ ($c_j^T$). The
basic variables of the map (Fig. \ref{map}) are then $c_i^T$ and
$c_j^T$ at time $t_n=nT$ of the $n^{th}+1$ stimulus, defined by
$x_n\equiv c_i^T(t_n)$ and $y_n\equiv (v_{\rm sr}/v_i)c_j^T(t_n)$
where both $x_n$ and $y_n$ are in units of $\mu$M, and the APD
corresponding to this stimulus, $A_{n+1}$.

The map is obtained by extending the restitution map to
include the effect of calcium on the APD and by
integrating the calcium flux equations
\begin{eqnarray}
\dot c_i^T&=&I_{\rm rel} -I_{\rm up}- \ica + \naca,\\
\dot c_j^T&=&(v_i/v_{\rm sr}) \left(-I_{\rm rel} + I_{\rm up}\right),
\end{eqnarray}
from time $t_n$ to time $t_{n+1}$. This yields
\begin{eqnarray}
A_{n+1}&=&F(D_n,x_n,y_n),\label{m1}\\
x_{n+1} &=& x_n + R_n-U_n+\Delta_n,\label{m2}\\
y_{n+1} &=& y_n - R_n+ U_n,\label{m3}
\end{eqnarray}
respectively, where
$R_n$, $U_n$, and $\Delta_n$ are the integrals
of $I_{\rm rel}$, $I_{\rm up}$, and $-\ica +\naca$
over the time interval $[t_n,t_{n+1}]$, respectively, and
are functions of $(D_n,x_n,y_n)$ for
a fixed pacing period; $v_iR_n$ and $v_iU_n$ are
the total amount of calcium released from and
pumped into the SR over one beat, respectively,
and $v_i\Delta_n$ is the net total calcium entry into the
cell over one beat which can be positive (negative) if the
exchanger extrudes more (less) calcium
from the cell than $\ica$ brings into the cell.

\begin{figure}
\includegraphics[height=5.0cm,width=5.0cm]
{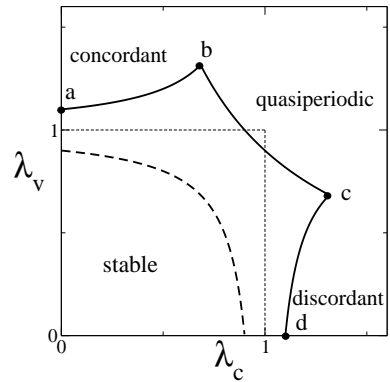} \caption{Stability boundaries
from the map analysis for positive
coupling $C=0.1$ with concordant alternans along
the dashed line, and negative coupling
$C=-0.1$ (solid line) with along the segments
$a-b$: concordant alternans; $c-d$: discordant alternans,
and $b-c$: quasiperiodicity.}
\label{stability}
\end{figure}

To study the stability of the fixed point of the map
$(A^*,x^*,y^*)$, we exploit the fact that the
total amount of calcium inside the cell is approximately
constant during steady-state pacing. Hence,
we can approximate the
3-dimensional (3-d) map (Eqs. \ref{m1}-\ref{m3}) by a
2-d map by assuming that $z_n\approx z^*$,
where $v_iz_n\equiv v_i(x_n+y_n)$ is the total
calcium in the cell at time $t_n$.
This 2-d map is given by
Eqs. \ref{m1} and \ref{m2} with $D_n=T-A_n$,
$\Delta_n=0$, and $y_n=z^*-x_n$.
A straightforward linear stability analysis of this
2-d map yields the eigenvalues
\begin{equation}
\lambda_{\pm}=\frac{1}{2}\left[-\lambda_v-\lambda_c\pm
\sqrt{(\lambda_c-\lambda_v)^2+4C}\right]
\end{equation}
where we have defined the quantities
\begin{eqnarray}
\lambda_v&=&\partial F/\partial D_n,\\
\lambda_c&=&-1-\frac{\partial (R_n-U_n)}{\partial x_n}+\frac{\partial 
(R_n-U_n)}{\partial y_n},\\
C&=&\frac{\partial (R_n-U_n)}{\partial D_n}\left(
\frac{\partial F}{\partial y_n}-\frac{\partial F}{\partial x_n}\right),
\end{eqnarray}
which are evaluated at the fixed point
of the map. Here, $\lambda_v$ and $\lambda_c$ govern the degree of
instability of the voltage and calcium systems, respectively,
while $C$ determines the sign of the coupling between the two
systems. Making APD-restitution ($\partial F/\partial D_n$) or the
relationship between release and SR-load ($\partial R_n/\partial
y_n$) steeper by increasing $\tau_f$ and $u$ in the ionic model is
equivalent to increasing $\lambda_v$ and $\lambda_c$,
respectively. Graded release
implies that $\partial (R_n-U_n)/\partial D_n$ is positive
for high pacing rates where $\ica$ depends on $D_n$,
such that the sign of $C$ is governed by
$\partial F/\partial y_n-\partial F/\partial x_n$ where the latter reflects
the effect of the magnitude of the calcium transient on APD via
$\ica$ and $\naca$ (Fig. \ref{coupling}). The periodic fixed point
undergoes a period doubling bifurcation when $|\lambda_-|=1$ and a
Hopf bifurcation for $(\lambda_v-\lambda_c)^2+4C<0$ when the pair
of complex eigenvalues $\lambda_\pm =re^{i(\pi\pm \omega)}$, with
$r=\sqrt{\lambda_c\lambda_v-C}$ and $\tan
\omega=\sqrt{-4C-(\lambda_c-\lambda_v)^2}/(\lambda_c+\lambda_v)$,
crosses the unit circle $(r=1)$. For the latter case, the
beat-to-beat oscillations of voltage and calcium are modulated
with a period $2\pi/\omega$. Examination of the eigenvectors for
$C<0$ reveals that alternans are discordant when $\lambda_-$ is
real and $\lambda_c>\lambda_v$. We plot in Fig. \ref{stability}
the corresponding stability boundaries for positive and negative
coupling in the $(\lambda_c,\lambda_v)$ plane which are remarkably
isomorphic to the stability boundaries obtained by simulations of
the ionic model in the $(u,\tau_f)$ plane of Fig. \ref{stab}. This
agreement shows that this simple map captures the main robust
features of the instability of the voltage-calcium system observed
in the ionic model and experimentally.

The numerical study of both the ionic model and the map in a
nonlinear regime reveals the existence of a rich dynamical
behavior including higher order periodicities  (3:3, 4:4, etc) as
well as transitions to chaos mediated by a period-doubling cascade
or intermittency depending on the parameters. Moreover, this model
naturally contains memory \cite{gilmour1,Watanabe}
due to the slow change of total calcium
concentration over several beats. Both of these aspects will be
discussed in more details elsewhere.

In conclusion, we have outlined the essential three-dimensional
parameter space that controls dynamic instability
of membrane voltage coupled to
calcium cycling and we have presented a theoretical framework in
which to interpret experiments beyond the limitations
of the one-dimensional restitution relationship.
The main axes of this parameter
space are the degree of instability of the voltage
and calcium systems, and the sign of the coupling
between the two systems, which is an important new parameter
to emerge from this work.  These results offer new concepts to
help identify the mechanisms which underly various heart rhythm
disorders. This research is supported by NIH SCOR P50-HL52319.

\end{document}